\documentclass[aps,floatfix,prd,nofootinbib,superscriptaddress,reprint,showpacs
  ,10pt,preprintnumbers,longbibliography]{revtex4-1}
\usepackage[utf8]{inputenc}
\usepackage[pdftex]{graphicx}
\usepackage{float}
\usepackage{amsmath}
\usepackage{amssymb}
\usepackage{mathtools}
\usepackage{braket}
\usepackage{siunitx}
\usepackage{amsfonts}
\usepackage{dsfont}
\usepackage{array}
\usepackage{bm}
\usepackage{mathrsfs}
\usepackage{pifont}
\usepackage{multirow}
\usepackage{upgreek}
\usepackage[dvipsnames]{xcolor}
\usepackage[pdftex,
  pdftitle={Digitised SU$(2)$ lattice gauge theory: interacting theory and
      Gauss invariance},
  pdfauthor={Timo Jakobs, Marco Garofalo, Johann Ostmeyer, Simone
      Romiti, Carsten Urbach},
  bookmarks,
  colorlinks,
  linkcolor=myblue,
  citecolor=mymagenta,
  menucolor=black,
  urlcolor=myblue,
  plainpages=false,
  pdfpagelabels,
  hypertexnames=false]{hyperref}
\usepackage{verbatim}
\usepackage{slashed}
\usepackage{cleveref}
\usepackage{ucs}
\usepackage{subcaption}

\usepackage{bbold}

\usepackage{tikz}
\usepackage{tikz-3dplot}
\usetikzlibrary{arrows}
\usetikzlibrary{decorations.markings}
\usetikzlibrary{calc}
\usetikzlibrary{decorations.fractals}
\usetikzlibrary{patterns,angles,quotes}
\usepackage{tikz-3dplot}

\DeclareMathOperator{\im}{i}

\definecolor{mymagenta}{RGB}{200, 0, 100}
\definecolor{myblue}{RGB}{45, 48, 146}

\graphicspath{{plots/}}

\begin{document}
\title{Dynamics in Hamiltonian Lattice Gauge Theory: Approaching the
  Continuum Limit with Partitionings of SU$(2)$}

\author{Timo Jakobs}
\affiliation{Helmholtz-Institut f\"ur Strahlen- und Kernphysik, University of
  Bonn, Nussallee 14-16, 53115 Bonn, Germany}
\affiliation{Bethe Center for Theoretical Physics, University of Bonn,
  Nussallee 12, 53115 Bonn, Germany}

\author{Marco Garofalo}
\affiliation{Helmholtz-Institut f\"ur Strahlen- und Kernphysik, University of
  Bonn, Nussallee 14-16, 53115 Bonn, Germany}
\affiliation{Bethe Center for Theoretical Physics, University of Bonn,
  Nussallee 12, 53115 Bonn, Germany}

\author{Tobias Hartung}
\affiliation{Northeastern University - London, Devon House, St Katharine Docks,
  London, E1W 1LP, United Kingdom}
\affiliation{Khoury College of Computer Sciences, Northeastern University,
  \#202, West Village Residence Complex H, 440 Huntington Ave, Boston, MA 02115,
  USA}

\author{Karl Jansen}
\affiliation{Computation-Based Science and Technology Research Center,
The Cyprus Institute, 20 Kavafi Street, 2121 Nicosia, Cyprus}
\affiliation{Deutsches Elektronen-Synchrotron DESY, Platanenallee 6, 15738
  Zeuthen, Germany}

\author{Johann Ostmeyer}
\affiliation{Helmholtz-Institut f\"ur Strahlen- und Kernphysik, University of
  Bonn, Nussallee 14-16, 53115 Bonn, Germany}
\affiliation{Bethe Center for Theoretical Physics, University of Bonn,
  Nussallee 12, 53115 Bonn, Germany}

\author{Simone Romiti}
\affiliation{Institute for Theoretical Physics, Albert Einstein Center for
  Fundamental Physics, University of Bern, CH-3012 Bern, Switzerland}

\author{Carsten Urbach}
\affiliation{Helmholtz-Institut f\"ur Strahlen- und Kernphysik, University of
  Bonn, Nussallee 14-16, 53115 Bonn, Germany}
\affiliation{Bethe Center for Theoretical Physics, University of Bonn,
  Nussallee 12, 53115 Bonn, Germany}
\date{\today}

\begin{abstract}
  In this paper, we investigate a digitised SU$(2)$ lattice gauge
  theory in the Hamiltonian formalism. We use partitionings to
  digitise the gauge degrees of freedom and show how to define a
  penalty term based on finite element methods to project onto physical
  states of the system. Moreover, we show for a single plaquette
  system that in this framework the limit $g\to0$ can be approached at
  constant cost.
\end{abstract}

\maketitle

\section{Introduction}
\label{sec:intro}

The implementation of SU$(N)$ lattice gauge theories in the
original formulation by Kogut and Susskind~\cite{Kogut:1974ag} is
notoriously difficult on both classical and quantum computers, at
least if one is interested in the limit of gauge coupling $g\to0$,
corresponding to the continuum limit of the lattice theory.
In combination with local gauge invariance, the non-Abelian structure
of such theories and the practical requirement for digitization and
truncation lead to non-localities in formulations suitable for
this limit, or severe increase in resource requirements.

While the widely used Clebsch-Gordan expansion~\cite{Zohar:2014qma} is
working well at large $g$, it is not well suited for the limit of
$g\to0$: the number of terms required in the expansion grows quickly
with decreasing values of $g$. The reason for this behaviour is likely
the fact that the electric part of the Hamiltonian is diagonal in this
formulation, which becomes less and less dominant in the foreseen
limit. 

Therefore, there is a significant effort to construct a basis in which
the magnetic part of the Hamiltonian is diagonal, which in general
involves some kind of gauge fixing and a suitable basis choice for the
gauge field degrees of freedom. Recently, in
Ref.~\cite{Grabowska:2024emw} a fully gauge fixed SU$(2)$ Hamiltonian
has been developed, based on ideas worked out in
Ref.~\cite{Bauer:2023jvw}. While the latter approach involves a
functional basis, the works in
Refs.~\cite{Gustafson:2022xdt,Gustafson:2023kvd} are based on discrete
tetrahedral and octahedral sub-groups of SU$(2)$.
Also, in Ref.~\cite{Burbano:2024uvn} a Gauge Loop-String-Hadron formulation
is developed on general graphs.
For earlier work see for instance
Refs.~\cite{Ji:2020kjk,Alexandru:2019nsa,Alexandru:2021jpm,Ji:2022qvr}.  
Of course, one can also try to find alternative Hamiltonians to the
one derived by Kogut and Susskind. Examples are quantum link
models~\cite{Chandrasekharan:1996ih,Brower:1997ha,Wiese:2021djl}, a
Hamiltonian based on a Heisenberg-Comb~\cite{Bhattacharya:2020gpm} or
the orbifold approach presented in \cite{Bergner2024}.

In Ref.~\cite{Jakobs:2023lpp} we have presented a formulation using
partitionings of SU$(2)$ based on Ref.~\cite{Hartung:2022hoz} (see
also Refs.~\cite{Jakobs:2022ugr,Garofalo:2023zkd}), which
has the advantage that the number of elements can be chosen freely
while working in the magnetic basis. In these references we have shown
how the canonical 
momenta and in particular their square can be constructed based on
finite element methods. We have tested this approach in the free
theory and found that the continuum energy levels and eigenstates are
recovered in the limit of continuous gauge symmetry. The disadvantage
of this approach is that gauge invariance is realised only
approximatively. For ways to mitigate this see Ref.~\cite{Romiti:2023hbd}.

In this paper we will use the same formalism and investigate its
behaviour in the interacting theory: for this, we show how to construct
the Gauss operator again based on finite element methods. This Gauss
operator can then be used to construct a penalty term, which lets one
single out the physically relevant states. We also introduce a
truncation which makes it possible to take the limit of gauge coupling
$g^2\to0$ (continuum limit) at constant cost and constant error stemming
from the group discretisation. This is exemplified for a single
plaquette system in the maximal tree gauge.

\section{Theory}

The Kogut and Susskind Hamiltonian~\cite{Kogut:1974ag} of lattice
gauge theory we alluded to in the previous section is defined on a
cubical lattice, discretising the spatial dimensions only. 
Similarly to Wilson's famous Lagrangian formulation of lattice gauge
theories~\cite{Wilson:1974sk}, the gauge degrees of freedom
take the form of links connecting the spatial lattice sites. Each
link is classically described by a colour matrix $U$ in the fundamental
representation of the gauge group $G$. 

Quantum states of the system are described by a wave function
\begin{equation}
    \psi \left(\left\{U_{\mathbf{x}, k}\right\} \right): G^{N_{\textrm{links}}}
    \, \,
    \rightarrow
    \, \, \mathbb{C} \, ,
\end{equation}
assigning a complex probability amplitude to every classical configuration
$\left\{U_{\mathbf{x}, k}\right\}$ of the gauge links. The indices
$\mathbf{x}$ and $k$ here label the location and direction of each link.

\subsection{Operators}

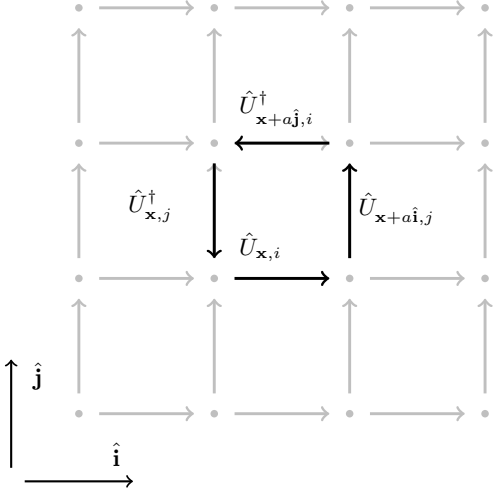
\begin{figure}[t]
    \begin{tikzpicture}
\draw[->,line width=0.8pt] (-0.7200000000000001,-0.9) -- (0.7200000000000001,-0.9);
\draw[->,line width=0.8pt] (-0.9,-0.7200000000000001) -- (-0.9,0.7200000000000001);
\node (e_mu) at (0.5,-0.54) {$\hat{\mathbf{i}}$};
\node (e_nu) at (-0.54, 0.5) {$\hat{\mathbf{j}}$};
\fill[color=gray!50] (0.0,0.0) circle(0.05);
\fill[color=gray!50] (0.0,1.8) circle(0.05);
\fill[color=gray!50] (0.0,3.6) circle(0.05);
\fill[color=gray!50] (0.0,5.4) circle(0.05);
\fill[color=gray!50] (1.8,0.0) circle(0.05);
\fill[color=gray!50] (1.8,1.8) circle(0.05);
\fill[color=gray!50] (1.8,3.6) circle(0.05);
\fill[color=gray!50] (1.8,5.4) circle(0.05);
\fill[color=gray!50] (3.6,0.0) circle(0.05);
\fill[color=gray!50] (3.6,1.8) circle(0.05);
\fill[color=gray!50] (3.6,3.6) circle(0.05);
\fill[color=gray!50] (3.6,5.4) circle(0.05);
\fill[color=gray!50] (5.4,0.0) circle(0.05);
\fill[color=gray!50] (5.4,1.8) circle(0.05);
\fill[color=gray!50] (5.4,3.6) circle(0.05);
\fill[color=gray!50] (5.4,5.4) circle(0.05);
\draw[->,color=gray!50,line width=1.1pt] (0.0,0.27) -- (0.0,1.53);
\draw[->,color=gray!50,line width=1.1pt] (0.27,0.0) -- (1.53,0.0);
\draw[->,color=gray!50,line width=1.1pt] (0.0,2.07) -- (0.0,3.33);
\draw[->,color=gray!50,line width=1.1pt] (0.27,1.8) -- (1.53,1.8);
\draw[->,color=gray!50,line width=1.1pt] (0.0,3.87) -- (0.0,5.13);
\draw[->,color=gray!50,line width=1.1pt] (0.27,3.6) -- (1.53,3.6);
\draw[->,color=gray!50,line width=1.1pt] (0.27,5.4) -- (1.53,5.4);
\draw[->,color=gray!50,line width=1.1pt] (1.8,0.27) -- (1.8,1.53);
\draw[->,color=gray!50,line width=1.1pt] (2.07,0.0) -- (3.33,0.0);
\draw[->,color=gray!50,line width=1.1pt] (1.8,2.07) -- (1.8,3.33);
\draw[->,color=gray!50,line width=1.1pt] (2.07,1.8) -- (3.33,1.8);
\draw[->,color=gray!50,line width=1.1pt] (1.8,3.87) -- (1.8,5.13);
\draw[->,color=gray!50,line width=1.1pt] (2.07,3.6) -- (3.33,3.6);
\draw[->,color=gray!50,line width=1.1pt] (2.07,5.4) -- (3.33,5.4);
\draw[->,color=gray!50,line width=1.1pt] (3.6,0.27) -- (3.6,1.53);
\draw[->,color=gray!50,line width=1.1pt] (3.87,0.0) -- (5.13,0.0);
\draw[->,color=gray!50,line width=1.1pt] (3.6,2.07) -- (3.6,3.33);
\draw[->,color=gray!50,line width=1.1pt] (3.87,1.8) -- (5.13,1.8);
\draw[->,color=gray!50,line width=1.1pt] (3.6,3.87) -- (3.6,5.13);
\draw[->,color=gray!50,line width=1.1pt] (3.87,3.6) -- (5.13,3.6);
\draw[->,color=gray!50,line width=1.1pt] (3.87,5.4) -- (5.13,5.4);
\draw[->,color=gray!50,line width=1.1pt] (5.4,0.27) -- (5.4,1.53);
\draw[->,color=gray!50,line width=1.1pt] (5.4,2.07) -- (5.4,3.33);
\draw[->,color=gray!50,line width=1.1pt] (5.4,3.87) -- (5.4,5.13);
\draw[->,line width=1.1pt] (2.07,1.8) -- (3.33,1.8);
\node[anchor=north west] (pp2) at (2.0160000000000005,2.52) {$\hat{U}_{\mathbf{x},i}$};
\draw[->,line width=1.1pt] (3.6,2.07) -- (3.6,3.33);
\node[anchor=north west] (pp2) at (3.6,3.06) {$\hat{U}_{\mathbf{x} + a\hat{\mathbf{i}},j}$};
\draw[->,line width=1.1pt] (3.33,3.6) -- (2.07,3.6);
\node[anchor=north west] (pp2) at (2.0160000000000005,4.41) {$\hat{U}_{\mathbf{x} + a \hat{\mathbf{j}},i}^\dagger$};
\draw[->,line width=1.1pt] (1.8,3.33) -- (1.8,2.07);
\node[anchor=north west] (pp2) at (0.54,3.096) {$\hat{U}_{\mathbf{x},j}^\dagger$};
    \end{tikzpicture}
    \caption{Sketch of the plaquette operator $\hat{P}_{\mathbf{x}, i j}$.}
    \label{fig:lattice_plaquette}
\end{figure}

To define the Kogut-Susskind Hamiltonian operators
$\hat{U}_{\mathbf{x}, k}$ are introduced, defined by the action 
\begin{equation}
    \hat{U}_{\mathbf{x}, k} \, \psi = U_{\mathbf{x}, k}\ 
    \psi\left(\dots, {U_{\mathbf{x}, k}}, \dots\right)
\end{equation}
on wave functions $\psi$, with $\hat{U}^\dagger \hat{U}^{~}= 1$ and
$\mathrm{det}\hat{U} =1$. 
Like the position operator in quantum mechanics the link operator modifies the
wave function by multiplying with the gauge link degree of freedom labelled
by $\mathbf{x}$ and $k$.
$\hat{U}$ and $\hat{U}^\dagger$ can then be combined to define the
plaquette operator $\hat{P}$. As depicted in 
\cref{fig:lattice_plaquette} it connects four gauge links to an oriented loop:
\begin{equation}
    \label{eq:plaquette}
    \hat{P}_{\mathbf{x}, i j} =\ \hat{U}_{\mathbf{x},i}\,
    \hat{U}_{\mathbf{x}+a\hat{\mathbf{i}}, j}\,
    \hat{U}^\dagger_{\mathbf{x}+a\hat{\mathbf{j}}, i}\,
    \hat{U}^\dagger_{\mathbf{x},j} \, .
\end{equation}
Furthermore, we define the left and right momentum operators
$\hat{L}^c_{\mathbf{x},k}$ and $\hat{R}^c_{\mathbf{x},k}$. They take the shape
of Lie derivatives and are defined as
\begin{align}
    \hat{L}^c_{\mathbf{x},k} \, \psi \	 & = -\im
    \frac{\mathrm{d}}{\mathrm{d}\beta}\,
    \psi \left(\dots, e^{- \im \beta t_c}\, U_{\mathbf{x},k},
    \dots\right)|_{\beta = 0}\,
    \intertext{and}
    \hat{R}^c_{\mathbf{x},k} \, \psi \	 & = -\im
    \frac{\mathrm{d}}{\mathrm{d}\beta}\,
    \psi \left(\dots,U_{\mathbf{x},k} \, e^{\im \beta
        t_c},\dots\right)|_{\beta = 0}\, ,
\end{align}
where the $t_c$ denote the generators of the gauge group.
The momentum operators obey the following
canonical commutation relations
\begin{align}
    [\hat{L}^c_{\mathbf{x},i}, \hat{U}_{\mathbf{y},j}]\  & =\
    -\delta_{\mathbf{x}
    \mathbf{y}} \, \delta_{ij} \, t_c\, \hat{U}_{\mathbf{x},i}\,, \\
    [\hat{R}^c_{\mathbf{x},i}, \hat{U}_{\mathbf{y},j}]\  & =\
    \delta_{\mathbf{x}
        \mathbf{y}} \, \delta_{ij} \, \hat{U}_{\mathbf{x},i} \, t_c \, ,
\end{align}
and
\begin{align}\label{eq:L}
    [\hat{L}^a_{\mathbf{x},i}, \hat{L}^b_{\mathbf{y},j}] & = \im f_{abc}
    \, \delta_{\mathbf{x}
        \mathbf{y}} \, \delta_{ij} \, \hat{L}^c \,\\
    \label{eq:R}
    [\hat{R}^a_{\mathbf{x},i}, \hat{R}^b_{\mathbf{y},j}] & = \im f_{abc}
    \, \delta_{\mathbf{x}
        \mathbf{y}} \, \delta_{ij} \, \hat{R}^c \, .
\end{align}
Here $f_{abc}$ are the structure constants of the gauge group.

\subsection{The Hamiltonian}

With these ingredients the Kogut-Susskind
Hamiltonian for a pure lattice gauge theory reads
\begin{equation}
    \label{eq:hamiltonian}
    \begin{split}
        \hat H &=\ \frac{g^2}{2}\sum_{\mathbf{x},c,k}
        \left(\hat{L}_{\mathbf{x},k}^c\right)^2 +
        \frac{2}{g^2}\sum_{\mathbf{x},j<i}
        \mathrm{Tr} \left[\,
        \mathbb{1} -
        \mathrm{Re}\,
        \hat{P}_{\mathbf{x}, ij} \right] \, .
    \end{split}
\end{equation}
The first term encodes the local kinetic energy and is typically referred to as
the electric
term. Its ground state is
\begin{equation}
    \psi_0^{\textrm{el.}} \left( \left\{ U_{\mathbf{x}, k} \right\}
    \right) = \mathrm{const} \, .
\end{equation}
The second term implements an interaction between the four links of each
plaquette and is typically referred to as the magnetic term. Its
ground state reads
\begin{equation}
    \psi_0^{\textrm{mag.}} \left( \left\{ U_{\mathbf{x}, k} \right\}
    \right) = \prod_{\mathbf{x}, i<j} \delta(\mathbb{1} -
    P_{\mathbf{x}, i j}) \, .
\end{equation}
The physical Hilbert space of the theory is further restricted by a constraint
referred to as Gauss law. It states that any physical state $\ket{\psi}$ needs
to satisfy
\begin{equation}
    \hat{G}^c_{\mathbf{x}} \ket{\psi} = \sum_{k} \left(
    \hat{L}^c_{\mathbf{x},k} + \hat{R}^c_{\mathbf{x} -
        a \hat{\mathbf{k}},k} \right) \ket{\psi} = 0 \, .
\end{equation}
This can be understood as demanding colour charge conservation at each
vertex in the lattice. It significantly reduces the dimensionality of the
Hilbert space of the theory.

\subsection{Dual Formulation}

By considering the ground states of the electric and magnetic parts of the
Hamiltonian alone, respectively, we can make some educated guesses
about the ground state of the full Hamiltonian. For large $g^2$ we
expect it to be quite uniform with little entanglement between the
links. This is because here we mostly have a free theory, perturbed by
a weak potential implemented by the magnetic term. 

When decreasing $g^2$ the entanglement between links increases with the
wave function only being non-vanishing for configurations where all the
$P_{\mathbf{x}, ij} \approx \mathbb{1}$. All other
configurations will be suppressed due to the then large $1/g^2$.
Thus, it would be highly beneficial to rewrite the Hamiltonian 
in terms of plaquette degrees of freedom instead of the original gauge links
\begin{equation}
    \psi \left(\left\{ U_{\mathbf{x}, i} \right\} \right) \rightarrow \psi
    \left( \left\{ P_{\mathbf{x}, ij} \right\} \right) \, ,
\end{equation}
which would lead to a magnetic term, consisting of a sum of single site
operators and nearest neighbour interactions in the electric term.
As a result, the entanglement between the individual degrees of
freedom would no longer increase for $g^2 \rightarrow 0$. Furthermore,
one could now make use of the fact that the wave function of the system
only is non-vanishing for $P_{\mathbf{x}, ij}$ close to the
identity. This could be exploited by choosing a basis for the wavefunctions 
that is suitable for approximating wavefunctions distributed
around $P_{\mathbf{x}, ij} = \mathbb{1}$ well.

While this idea can be implemented in an Abelian U$(1)$ theory, it is
obfuscated in non-Abelian theories by their non-commutative nature,
which makes it necessary to add additional terms to the Hamiltonian,
which introduce non-localities.
For SU$(N)$ multiple dual Hamiltonians are  under consideration
\cite{Bauer:2023jvw,LIGTERINK2000215,PhysRevD.107.074504}.

Since this is not the focus of this paper, we avoid this complication
by studying a single plaquette lattice only.
By squaring Gauss' law it is easy to show that
\begin{equation}
    \sum_c (\hat{L}_1^c)^2 = \sum_c (\hat{L}_2^c)^2 = \sum_c (\hat{L}_3^c)^2 =
    \sum_c (\hat{L}_4^c)^2\,,
\end{equation}
where the indices 1,2,3 and 4 label the four links in the plaquette. Thus, one
can express the Hamiltonian in terms of a single gauge degree of
freedom (equivalent to the single plaquette operator)
\begin{equation}
  \hat{H} = 2 g^2 \sum_c (\hat{L}^c)^2 + \frac{2}{g^2}\mathrm{Tr} \left[
    \mathbb{1} - \hat{U} \right]\,.
\end{equation}
The remaining Gauss law constraint at the origin then takes the shape of
\begin{equation}
    \hat{Q}^c \ket{\psi} = (\hat{L}^c + \hat{R}^c) \ket{\psi} = 0 \,.
\end{equation}
In the following we will enforce Gauss' law by adding a penalty term
\begin{equation}
    \hat{H}_{\textrm{penalty}} \ = \ \kappa \sum_{c} \left( \hat{Q}^c \right)^2
    \label{eq:spPenaltyTerm}
\end{equation}
to the Hamiltonian of the theory. Here, $\kappa$ is a large positive
constant. Such a penalty term will shift unphysical states to higher
energies\cite{PhysRevLett.125.030503}, allowing for simulations of the low-lying 
physical spectrum of the theory.

Furthermore, it is analytically known~\cite{Bauer:2023jvw} that the
physical states are the ground state and the fourth excited state of
the unconstrained Hamiltonian. Therefore, this system also allows us to
study the practicability of using a penalty term.

The gauge group of interest in the following is SU$(2)$. Thus, the generators
are given by $t_c = \frac{1}{2} \sigma_c$, where $\sigma_c$ are the Pauli
matrices. The structure constants $f_{abc}$ are the components of the Levi-Civita
tensor. SU$(2)$ serves as a useful toy model to explore the challenges
surrounding Hamiltonian simulations with non-abelian gauge groups. The lessons
learned here should hopefully lead the way to Hamiltonian simulations of
quantum chromodynamics.

\section{Discretising the Operators}
\label{sec:discretisation}

As the Hilbert space of wave functions is in principle infinite-dimensional,
in general a discretisation and possibly a truncation is needed for a
practical numerical simulation. The
full Hilbert space of the theory can be decomposed into products of wave
functions on single gauge links
\begin{equation}
  \ket{\{f_{\mathbf{x},k}\}} = \bigotimes_{\mathbf{x},k}
  \ket{f_{\mathbf{x},k}}\,.
\end{equation}
Thus, it is sufficient to find discretisation schemes for the single link wave
functions
\begin{equation}
  f_{\mathbf{x},k}(U) : G \rightarrow \mathbb{C} \, .
\end{equation}
More specifically, we will use the finite element canonical momenta we
presented in Ref.~\cite{Jakobs:2023lpp}. The idea here is to
approximate each link wave function at a finite set of gauge group
elements
\begin{equation}
  \left\{ D_i \right\} \subset \mathrm{SU}(2) \, .
\end{equation}
In the following we will refer to such a subset as a partitioning of SU$(2)$.
Any such partitioning can be connected to a simplical mesh $\{(i_0, i_1, i_2,
  i_3)\}$ via a Delaunay triangulation~\cite{Delaunay:1934}.
Here, the integers $i_j$ label the four group elements spanning each
simplex in the mesh.

Next one can introduce the basis functions $\hat{\phi}_i$ with the
property
\begin{equation}
  \hat{\phi}_i (D_j) = \delta_{ij}\,,
\end{equation}
and interpolate linearly inside each simplex of the mesh. Within each simplex
we introduce local coordinates $\vec{\alpha}_{L/R}$ defined by
\begin{align}
  \label{eq:locLeftCoords}
  U = \exp\left( -\im \vec{\alpha}_{L} \cdot \vec{t} \, \right) D_{i_0}
  \intertext{and}
  \label{eq:locRightCoords}
  U = D_{i_0} \exp\left( \im \vec{\alpha}_{R} \cdot \vec{t} \, \right)\,,
\end{align}
respectively. The
local coordinates are chosen such that the left and right
canonical momentum operators take the shape of the components of Lie
derivatives on $S_3$. By Taylor expanding the function around the value at
each vertex, one can calculate the Lie derivatives within each cell to be
defined by
\begin{equation}\label{eq:derivative_FVE}
  \begin{pmatrix}
    \vec{\alpha}_{1,L/R}^T \\
    \vec{\alpha}_{2,L/R}^T \\
    \vec{\alpha}_{3,L/R}^T
  \end{pmatrix} \vec\nabla_{C,L/R} \, f= \begin{pmatrix}
    f(D_{i_1}) - f(D_{i_0}) \\
    f(D_{i_2}) - f(D_{i_0}) \\
    f(D_{i_3}) - f(D_{i_0})
  \end{pmatrix}\,,
\end{equation}
where $\vec{\alpha}^T_j$ denotes the coordinates of the vertices $D_{i_j}$
found in the simplex $C$. To then further improve the estimate of the momentum
operators at a given vertex, one can average the Lie derivative over the
simplices surrounding a given vertex. In our implementation this average is
weighted by each cell's volume. The operator matrices are thus calculated as
\begin{equation}
  \begin{pmatrix}
    \hat{L}^1_{ij} \\
    \hat{L}^2_{ij} \\
    \hat{L}^3_{ij}
  \end{pmatrix} = \frac{- \im}{\sum_{\{C|i \in C\}} \mathrm{Vol(C)}}
  \sum_{\{C|i \in C\}} \mathrm{Vol(C)} \, \, \vec\nabla_{C,L}
  \hat{\phi}_j \,.
\end{equation}
Using the $L^a$ operators obtained above to construct the Laplacian operator
$\sum_c  (\hat{L}^c)^2$
will result in a poor approximation because the $L^a$ construction relies on a
linear approximation.
A direct construction of the Laplacian operator
$\sum_c (\hat{L}^c)^2$ can be obtained as in \cite{Jakobs:2023lpp}
\begin{equation}
  \sum_{c}
  \left(\hat{L}^c\right)^2 = \frac{1}{v(i)} \sum_{\{C
    \in \mathcal C| i,j \in C\}}(\vec\nabla_C \hat{\phi}_i) \cdot
  (\vec\nabla_C \hat{\phi}_j) \, \mathrm{Vol} (C)\,.
  \label{eq:L_Greens}
\end{equation}
where $v(i)$ here denotes the barycentric weight at vertex $i$. They are
obtained by
equally distributing the volume of each simplex onto its vertices
\begin{equation}
  v(i) = \frac{1}{4} \sum_{\{C | i \in C\}} \mathrm{Vol}(C) \, .
\end{equation}
Again, using the $\hat{L}$ and $\hat{R}$ operators na\"ively to
construct the squared Gauss operator needed for the penalty term leads
to a large artefacts. However, similarly to the Laplacian, also the
squared Gauss operator
\begin{equation}
  \sum_c \hat{Q}_c^2 =	\sum_c \left(\hat{L}_c + \hat{R}_c \right)^2 \, .
\end{equation}
can be constructed directly. One obtains
\begin{equation}
  \begin{split}
    \sum_{c}
    \left(\hat{Q}^c\right)^2 &= \frac{1}{v(i)}	\sum_{\{C
      \in \mathcal C| i,j \in C\}} \, \mathrm{Vol} (C) \\
    & ((\vec\nabla_{C,L} + \vec\nabla_{C,R}) \hat{\phi}_i) \cdot
    ((\vec\nabla_{C,L} + \vec\nabla_{C,R}) \hat{\phi}_j) \,.
  \end{split}
  \label{eq:Q_Greens}
\end{equation}
Here $\vec\nabla_{C,L}$($\vec\nabla_{C,R}$) denote the cell gradient taken in
the left (right) local coordinates.

All these momentum operators are local in the sense that a given vertex is
only ever mapped to vertices it shares a simplex with. Furthermore, the gauge
link operator $\hat{U}$ is diagonal in this basis.
However, for these operators the canonical commutation relations and the
low-lying spectrum are only exactly recovered in the limit of infinitely fine
meshes.
For a finite set of gauge group elements the canonical commutation relation are
only approximatively fulfilled.

\subsection{SU$(2)$ Partitioning}

\begin{figure*}
  \centering
  \begin{subfigure}[c]{0.4\textwidth}
    \includegraphics[width=0.95\columnwidth]{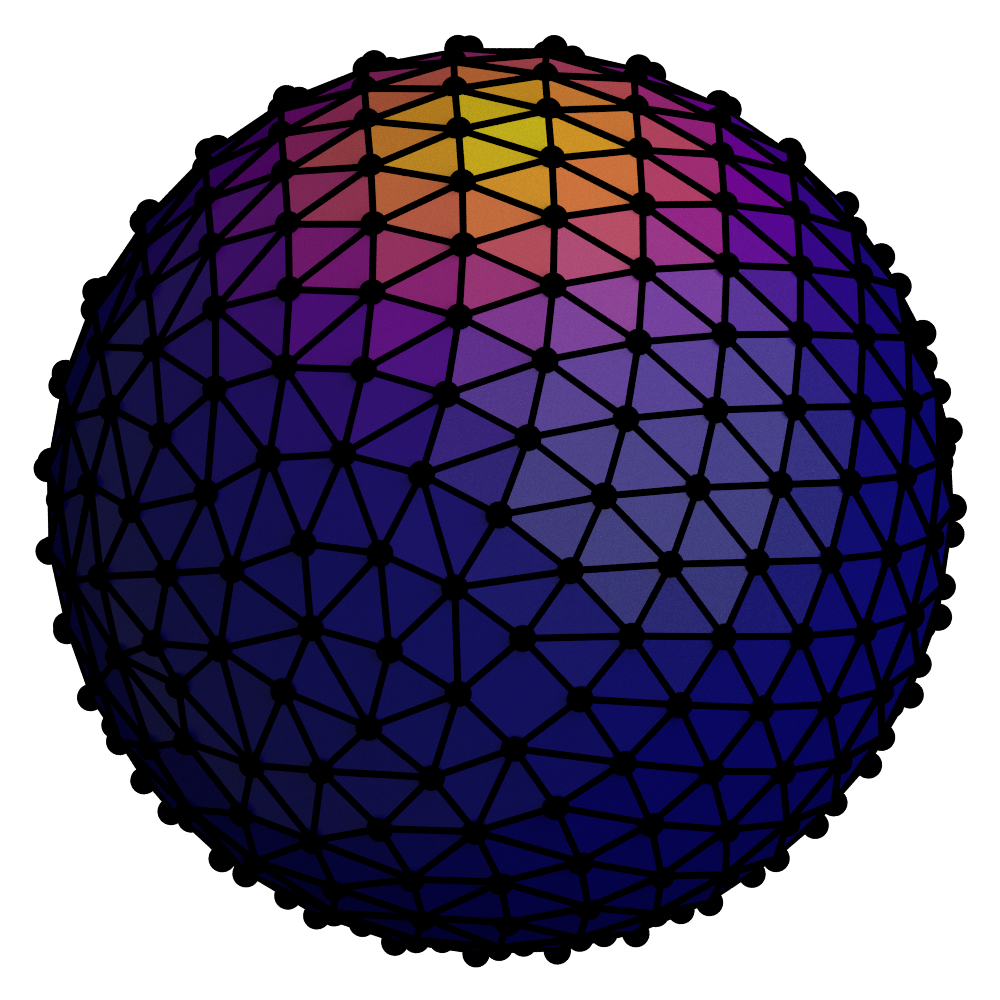}
  \end{subfigure}
  \begin{subfigure}[c]{0.15\textwidth}
    \scalebox{3}{$\rightarrow$}
  \end{subfigure}
  \begin{subfigure}[c]{0.4\textwidth}
    \includegraphics[width=0.95\columnwidth]{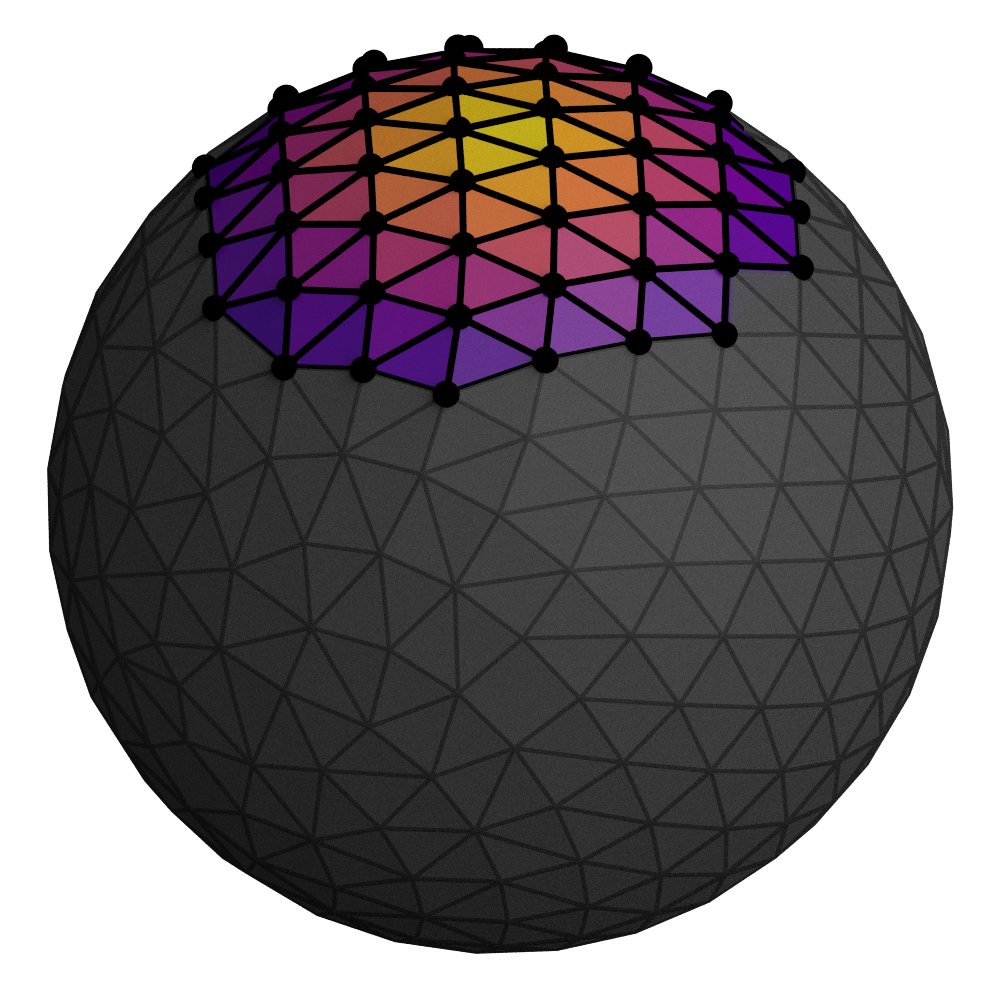}
  \end{subfigure}
  \caption{Sketch of the truncation procedure used in the dual Hamiltonian: On
    the left sphere we show the expected shape of the plaquette wave function
    towards weaker coupling. Warmer colours symbolise bigger probability
    amplitudes in the wavefunction. By using partitionings fulfilling
    \cref{eq:partitioningTruncation}, we effectively restrict our simulations to
    a
    cone surrounding the north-pole of the sphere. This is shown on the right.
    Note that SU(2) is isomorphic to the 3-sphere embedded in four dimensions,
    i.e.\ the
    pictures shown here should be understood as a lower dimensional sketch of
    the procedure.}
  \label{fig:poleTruncation}
\end{figure*}

In the following we will use the rotated simple cubic (RSC)
partitionings. These are obtained by constructing a simple cubical lattice in
the unit cube:
\begin{equation}
  \Lambda_n^{\textrm{SC}} = \left\{\vec x\in [0,1)^3 \,\middle|\, \vec x =
  d_{\textrm{SC}}(n)\, \textrm{R}\, \vec m + \vec a
  \,, \vec m \in \mathbb{Z}^3 \right\} \, .
\end{equation}
Here $n$ denotes the target number of points in the lattice.
$d_{\textrm{SC}}(n)$
is the distance between neighbouring points given by
\begin{equation}
  d_{\textrm{SC}}(n) = n^{-1 / 3} \, .
\end{equation}
The rotation matrix R is needed to ensure that the planes of the lattice are
not aligned with the faces of the unit cube. In our implementation successive
rotations of $\pi/8$ around $\hat{e}_1$,$\hat{e}_2$ and $\hat{e}_3$ seem to
work well. The vector $\vec a$ is tuned such that the number of points actually
in the cube equals the target $n$.

This lattice is then mapped SU(2) via the volume preserving map:
\begin{equation}
  \begin{split}
    \rho(x_1) & =  \Phi_1 (x_1) \,,\\
    \theta(x_2) & =  \cos^{-1}\left(1-2 \,x_2\right)\\
    \text{and} \qquad \phi(x_3) & = 2 \pi x_3 \, .
  \end{split}
  \label{eq:uniform-s3-map}
\end{equation}
Here the function $\Phi_1(\rho)$ is defined via its inverse
\begin{equation}
  \Phi_1^{-1} (\rho) = \frac{1}{\pi}  \left( \rho - \frac{1}{2} \sin( 2 \rho)
  \right)\,.
\end{equation}
The angles $(\rho, \theta, \phi)$ parametrise an SU(2) elements by
\begin{equation}
  U(\rho, \theta, \phi) = \cos \rho \mathbb{1} + \im \sin \rho \,
  \vec{e}_{\rho}(\theta, \phi) \cdot \vec{\sigma}\,,
\end{equation}
where $\vec{e}_{\rho}(\theta, \phi) = (\sin\theta \cos \phi, \sin
  \theta \sin \phi, \cos \theta)$ is a point on $S_2$.
For more details we refer to \cite{Jakobs:2023lpp,Hartung:2022hoz}. \\

Moreover, as depicted in \cref{fig:poleTruncation}, we expect the
plaquette wave function of low-lying states at weak couplings to
vanish for points far away from $\mathbb{1}$. For this purpose we
implement a truncation by  modifying the volume preserving map such
that
\begin{equation}
  \label{eq:partitioningTruncation}
  \mathrm{Tr} \left[	D_i \right] \geq 2 \cos \left( \pi
  \varepsilon_{\mathrm{Tr}}\right) \, .
\end{equation}
Here the parameter $\varepsilon_{\mathrm{Tr}} \in (0,1]$ controls how
much of the gauge group is approximated by the partitioning around
the identity. $\varepsilon_{\mathrm{Tr}}
  = 1$ gives points covering the entire group, while
$\varepsilon_{\mathrm{Tr}}=0$ only allows for $D_i = \mathbb{1}$. \\

In the following we will denote different partitionings as
RSC$_N^{\varepsilon_\mathrm{Tr}}$ describing a set of $N$ points found
within the hypersphere cap defined by \cref{eq:partitioningTruncation}.

c\section{Numerical Results}\label{sec:results}

To study the performance of the proposed discretisation scheme, the ground
state and first excited state for the single plaquette system are determined
via exact diagonalisation. Gauss' law will be enforced either by manual
selection of the correct states or by the penalty term
\cref{eq:spPenaltyTerm}.

As observables, we will consider the ground state energy $E_0$, the
mass gap $M$ defined as
\begin{equation}
  M = E_1 - E_0
\end{equation}
and the ground state plaquette expectation value
\begin{equation}
  \langle P \rangle \ =\ \frac{1}{2 \, N_{\textrm{Plaq}}} \,
  \sum_{\mathbf{x}, i<j} \mathrm{Tr} \left[
  \bra{\psi_0} \hat{P}_{\mathbf{x}, ij} \ket{\psi_0} \right] \, .
  \label{eq:pExpValue}
\end{equation}
Our numerical results can be compared to the analytic solution derived in
Ref.~\cite{Ligterink:2000ug}. There the energy levels are calculated to be
\begin{equation}
  E_n = \frac{4}{g^2} + \frac{g^2}{8} \left(b_{2n}\left(-16/ g^4 \right) -
  4 \right) \,.
\end{equation}
Here $b_n$ denotes the Mathieu characteristic numbers. Note that some of the
prefactors differ from the cited source due to a difference in convention
used for the Hamiltonian. The theoretical prediction of the plaquette
expectation value is obtained by numerically integrating the eigenfunctions
also given in Ref.~\cite{Ligterink:2000ug}.

\subsection{Tuning the Penalty Term}

For simulations with a penalty term \cref{eq:spPenaltyTerm}, a value
for the parameter $\kappa$ needs to be chosen. For formulations with
only approximative gauge invariance, this can be delicate. Since
$\hat{Q}^c$ only approximates the exact Gauss operator,
$\hat{Q}^c|\psi\rangle$ can be non-zero even for a physical wave
function $|\psi\rangle$. While this is expected to be a small effect
in $\hat{Q}^c|\psi\rangle$, large values of $\kappa$ can inflate
it. Thus, $\kappa$ should not be chosen much larger than needed to
move unphysical states beyond the energies of interest. Otherwise, the ground
state of the total system will simply be the one with the most favourable
discretisation error.

Furthermore, much better matching with the analytic predictions can be
achieved, when correcting for energy contributions by the penalty term. This is
done by simply subtracting the expectation value of the penalty term from the
eigenvalues obtained from the solver:
\begin{equation}
  E_i = \lambda_i - \langle H_{\textrm{penalty}} \rangle\,.
  \label{eq:penaltycorrection}
\end{equation}
One way to test whether $\kappa$ is sufficiently large is to study the
expectation value of $\hat{H}_{\textrm{penalty}}$. This shows a sharp drop,
once the unphysical states are projected out.

\subsection{Overview}

\begin{figure*}
  \begin{subfigure}[t]{0.48\textwidth}
    \includegraphics[width=0.95\columnwidth]{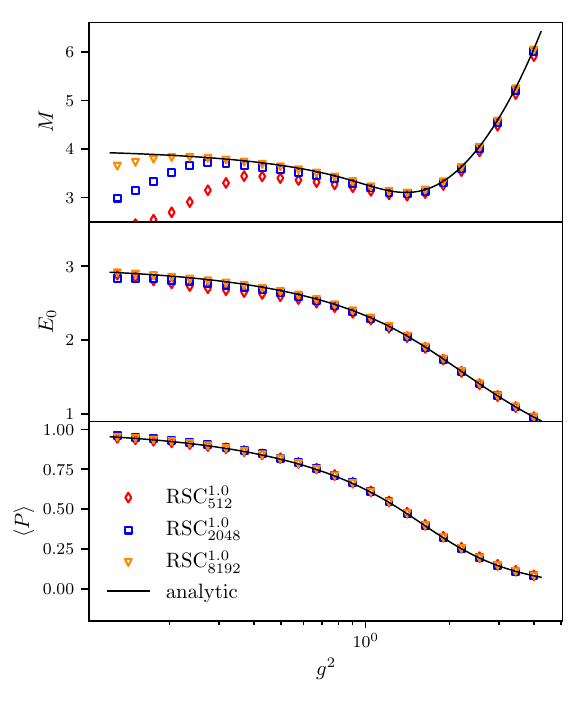}
    \caption{Gauss law implemented via manual state selection.}
    \label{fig:mtOverviewNoPenalty}
  \end{subfigure}
  \hfill
  \begin{subfigure}[t]{0.48\textwidth}
    \includegraphics[width=0.95\columnwidth]{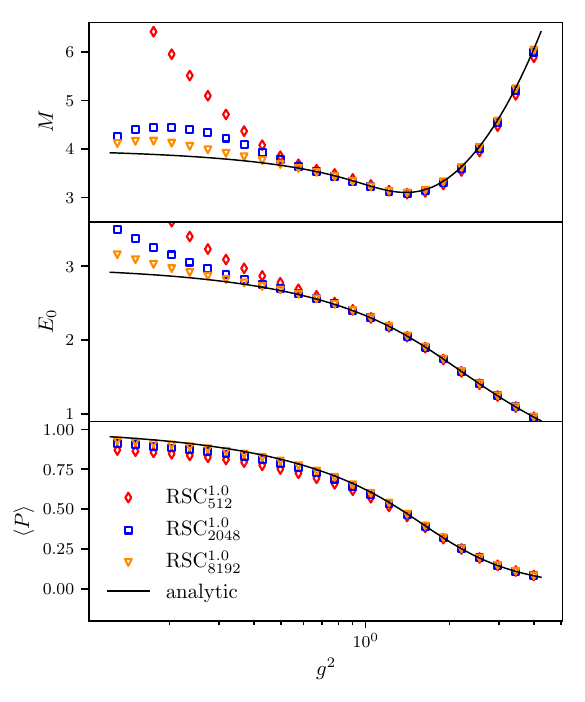}
    \caption{Gauss law implemented via penalty term with $\kappa=5$.}
    \label{fig:mtOverviewPenalty}
  \end{subfigure}
  \caption{Shown are the mass gap $M$, the ground state energy $E_0$ and the
    ground state plaquette expectation value $\langle P \rangle$ (as
    defined in
    \cref{eq:pExpValue}) as a function of the coupling $g^2$ for three
    different
    partitioning sizes. On the left physical states are selected manually
    while on the right a penalty term is used. The solid line shows the
    analytic prediction of each observable.}
  \label{fig:mtOverview}
\end{figure*}

In \cref{fig:mtOverview} we show all three observables as a
function of the squared coupling $g^2$. For \cref{fig:mtOverviewNoPenalty},
physical states were selected manually -- which is possible since the
analytic solution is known, while in \cref{fig:mtOverviewPenalty}
the penalty term with $\kappa=5$ was used. The data points are
obtained for partitionings with $256$ (red diamonds), $1024$ (blue
squares) and $4096$ (orange triangles) elements and cover the whole
gauge group without truncation (corresponding to $\varepsilon_\mathrm{Tr}=1$).

At larger values of the coupling, we see good agreement of all
observables with the corresponding analytic prediction represented by
the black, continuous lines.
Towards smaller couplings, however, the simulation results
increasingly deviate. As expected, these deviations are
biggest for the coarse partitioning and decrease when going to finer
partitionings.

Curiously, the amount and sign of the deviation changes, depending on
whether physical states are selected manually or via the penalty
term. For the former the mass gap and ground state energy are
underestimated towards small $g^2$. The penalty term however, leads to
an overestimation of the ground state energy and mass gap, while the
plaquette expectation value is notably smaller than predicted.
Note that the energies as plotted are already corrected for the penalty term
expectation value as described in \cref{eq:penaltycorrection}.

To explain this we can take a look at the plaquette expectation value. For the
manual state selection the analytic prediction is matched well, even for very
small $g^2$. This suggests, that the correct states are still found, but their
electric energy is underestimated by the meshed operators.

When using the penalty term however, the measured plaquette expectation values
are found to be smaller than the prediction. This is likely because states with
a large plaquette expectation value have more electrical energy and thus also
lead to larger discretisation errors in the penalty term.

Lastly we note, that deviations appear to be largest in the mass gap.
Thus, we will focus on this observable for our remaining tests.

\begin{figure}
  \centering
  \includegraphics[width=0.95\columnwidth]{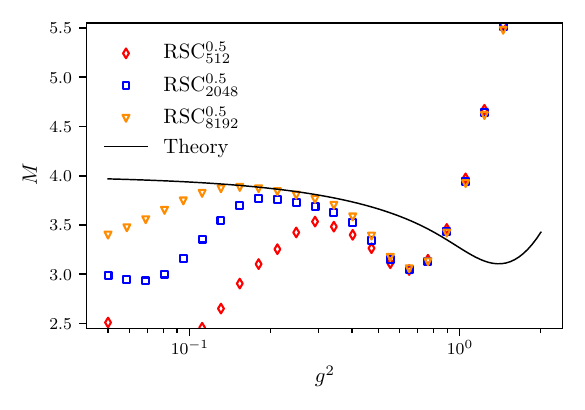}
  \caption{Plotted is the mass gap $M$ as a function of the coupling $g^2$
    for three partitioning sizes with $\varepsilon_{\mathrm{Tr}} = 0.5$.
    The analytic prediction is again shown in black.}
  \label{fig:mtTruncOverview}
\end{figure}

\subsection{Truncation}

\begin{figure}
  \centering
  \includegraphics[width=0.95\columnwidth]{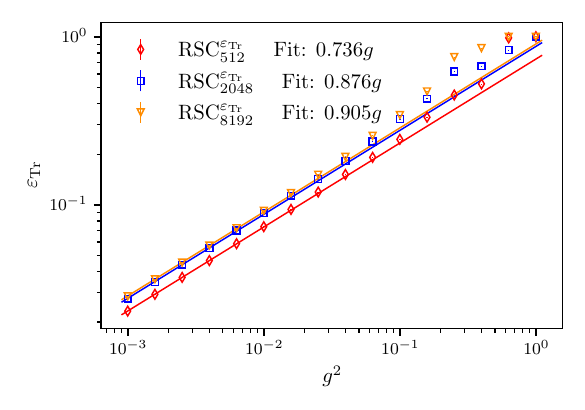}
  \caption{Plotted are the optimal values for $\varepsilon_{\mathrm{Tr}}$ as
    a function of the coupling for three different partitionings. The optimal
    value is found by minimizing the deviation from the analytic prediction
    of the mass gap. The error bars show the resolution of the scan used to
    find this minimum. Also shown are fits to 5 leftmost points in the plot.}
  \label{fig:mtEpsTrCalib}
\end{figure}

Next we would like to test partitionings with $\varepsilon_\mathrm{Tr}<1$.
We expect an interval $I$ in $g^2$ where the approximation works best: At large
$g^2$-values errors due to too small $\varepsilon_\mathrm{Tr}$ will dominate,
while at small values of $g^2$, the resolution around the identity is
insufficient.

Moreover, increasing the resolution of the partitioning at fixed
$\varepsilon_\mathrm{Tr}$ should move this interval to smaller values of $g^2$
and decrease the overall deviation in this interval region to the exact result.

This expectation is confirmed by our simulations. As an example we
show in \cref{fig:mtTruncOverview} the mass gap as a function of the coupling
for three different partitioning sizes with $\varepsilon_\mathrm{Tr}=0.5$. As
predicted there is a matching interval at $g^2 \approx 0.2$, which increases in
size and moves to smaller couplings with finer partitionings.

In practice, this means one would rather choose $\varepsilon_{\mathrm{Tr}}$ too
large than to small, as the former will still recover the correct result in the
limit of infinitely fine partitionings, be it at higher cost.

In order to determine an appropriate truncation for a given coupling,
we determined $\varepsilon_{\mathrm{Tr}}^\mathrm{opt}$ such that the
deviation from the analytic prediction of the mass gap is minimized. The
results can be found in \cref{fig:mtEpsTrCalib}, where the such determined
$\varepsilon_{\mathrm{Tr}}^\mathrm{opt}$ is plotted as a function of $g^2$,
again for three different partitioning sizes. For about $g^2<0.1$ the
dependence is to a good approximation linear in $g$.
This is in agreement with the results in Ref.~\cite{Bauer:2023jvw}.
$\varepsilon_{\mathrm{Tr}}^\mathrm{opt}$ also increases with partitioning size.
This is the same effect mentioned earlier: the matching interval moves to
smaller coupling values with increasing partitioning size at fixed
$\varepsilon_{\mathrm{Tr}}^\mathrm{opt}$.

To predict the optimal truncation for a given coupling, we performed
linear fits in $g$ to the data. As we are most interested in small
couplings, only the five leftmost points in the plot were used for the
fit. The truncation 
parameters derived from these fits will be referred to as
$\varepsilon_\mathrm{opt}$ in the following.

\begin{figure}
  \includegraphics[width=0.95\columnwidth]{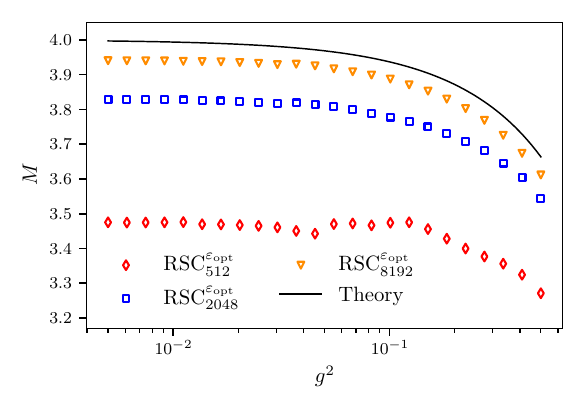}
  \caption{The mass gap $M$ as a function of the coupling $g^2$ for different
    fixed partitioning sizes. $\varepsilon_{\mathrm{Tr}}$ is chosen
    separately at each coupling according the fits found in
    \cref{fig:mtEpsTrCalib}.}
  \label{fig:mtOptiEps}
\end{figure}

Finally, in \cref{fig:mtOptiEps} we show the mass gap $M$ as a
function of $g^2$, again for three resolutions simulated with the
corresponding $\varepsilon_{\mathrm{Tr}}^\mathrm{opt}(g^2)$
and compared to the analytical results. The finer the resolution, the
smaller the deviation from the analytical curve at each
$g^2$-value. And, the deviation for each partitioning size appears to
become independent of $g^2$ in the $g\to0$ limit.

\subsection{Convergence and Cost}
\label{subsec:convergence}

\begin{figure}
  \includegraphics[width=0.95\columnwidth]{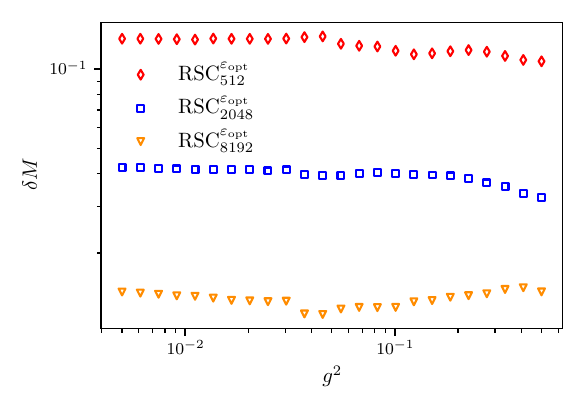}
  \caption{The relative deviation $\delta M$ of the mass gap shown in
    fig.~\ref{fig:mtOptiEps}
    as a function of $g^2$ at fixed partitioning sizes.}
  \label{fig:mtOptiEpsRel}
\end{figure}

\begin{figure}
  \includegraphics[width=0.95\columnwidth]{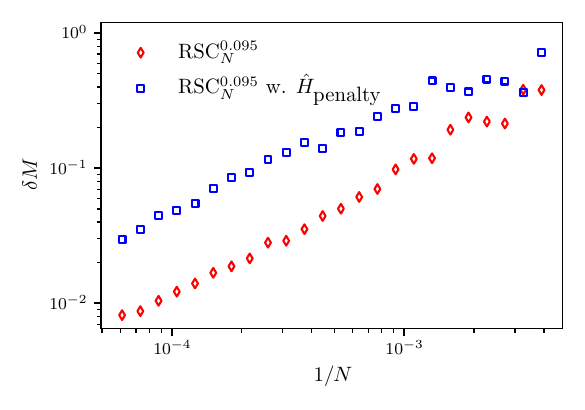}
  \caption{The relative deviation of the mass gap $\delta M$ as a function of
    the inverse partitioning size $1/N$ at a coupling of $g^2 = 0.01$ and
    $\varepsilon_{\mathrm{Tr}} = 0.92 g $.
    Results are obtained with penalty term (blue) and without (red).}
  \label{fig:mtConvergence}
\end{figure}

To get a more quantitative idea of this and the accuracy of the
partitionings, we study the relative deviation
\begin{equation}
  \delta M = \frac{\left|M -
    M_{\mathrm{analytic}}\right|}{M_{\mathrm{analytic}}} \, .
\end{equation}
$\delta M$ is plotted as a function of $g^2$ in
\cref{fig:mtOptiEpsRel}. There one can observe more clearly that the
deviations are mostly independent of the coupling for $g^2  \to 0$,
which means the weak coupling limit can be approached at approximately
constant simulation cost and uncertainty. 
Next, we plot the relative deviation as a function of the
inverse partitioning size in \cref{fig:mtConvergence} at $g^2=0.01$.
Here, the red diamonds are obtained by manual selection of the physical
states from the Hamiltonian, while the blue points are obtained using
the penalty term with $\kappa=5$. The results obtained with the penalty
term show larger deviations, but both approaches seem to have a similar
convergence rate. While selecting the physical state manually seems to produce
more accurate results, it becomes quickly unfeasible when the system size
increases.

\subsection{Extrapolating to the Full Group}
\label{sec:extrapolation}

\begin{figure*}
  \begin{subfigure}[t]{0.48\textwidth}
    \includegraphics[width=0.95\columnwidth]{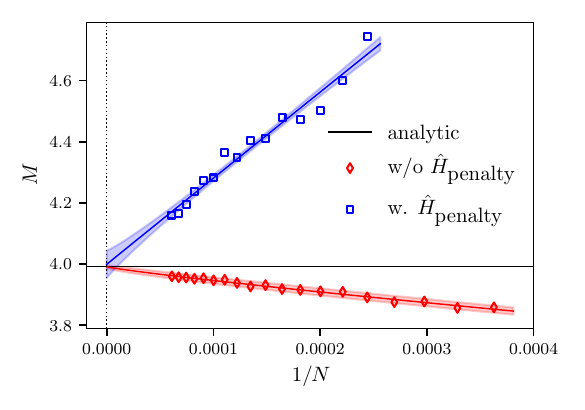}
    \caption{Plotted is the mass gap $M$ as a function of the inverse
    partitioning size at a coupling of $g^2=0.01341$. Results are
    obtained with penalty term (blue) and without (red). In black,
    we plot the analytic solution. Also included are fits according
    to \cref{eq:fitInN}. These allow us to extrapolate to
    $N \to \infty$.}
    \label{fig:mtGrpExtr}
  \end{subfigure}
  \hfill
  \begin{subfigure}[t]{0.48\textwidth}
    \includegraphics[width=0.95\columnwidth]{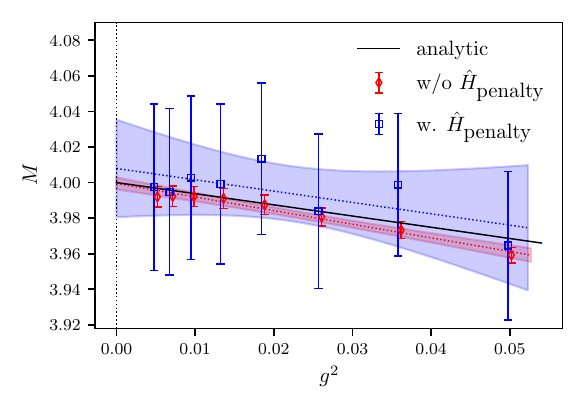}
    \caption{Plotted is the mass gap in the limit $N \to \infty$ as a 
    function of the coupling $g^2$. The data is obtained from fits like
    the one shown on the left. The analytic solution is again plotted
    in black. Also included are linear fits to the data. These allow us
    to extrapolate to $g^2 \to 0$.}
    \label{fig:mtContAfterGrp}
  \end{subfigure}
  \caption{These two figures contain extrapolations for the mass gap $M$
  first in the limit of $N\to\infty$ (left) and then $g^2 \to 0$ (right). This
  is done to determine the mass gap one would expect for the full gauge
  group in the continuum limit. The truncation parameter is chosen to be
  $\varepsilon_{\mathrm{Tr}} = 1.1 g$ when the penalty term is used, 
  and $\varepsilon_{\mathrm{Tr}} = 0.95g$ without.}
  \label{fig:mtContinuumAfterGroupLimit}
\end{figure*}
Lastly, we would like to test whether we can successfully extrapolate first to
the full gauge group and then to $g^2=0$. To do so we collect data at eight
different couplings {$0.005 \leq g^2 \leq 0.05$} and partitioning sizes
up to {$N = 2^{14} = 16\,384$}. When using the penalty term with $\kappa=5$,
${\varepsilon_{\mathrm{Tr}} = 1.1 g}$ was chosen, for the manual state selection
${\varepsilon_{\mathrm{Tr}} = 0.95 g}$. Choosing $\varepsilon$ a bit larger, when
using the penalty term appears to improve results significantly, which
at this point is a purely empirical observation.

In the first step we then perform a least squares fit to extrapolate to
infinitely fine partitionings at fixed coupling. The Ansatz for the fit
function reads
\begin{equation}
  M_{g^2}(N) = M^0_{g^2} \pm \left(\frac{N^0_{g^2}}{N}\right)^b \,.
  \label{eq:fitInN}
\end{equation}
Here $M^0_{g^2}$ and $N^0_{g^2}$ are treated as separate parameters for each
coupling, while the exponent $b$ is fitted globally to all eight datasets for
different couplings. The sign is chosen positive for manual state selection
and negative for the penalty term data.

This ansatz appears to describe the data well, while the global fit
produces more stable results as compared to separate fits per $g$-value.
As an example, we show the resulting fit for $g^2=0.01341$ together
with the data in \cref{fig:mtGrpExtr}. The (global) convergence rate
is measured to be $b = 1.00(11)$ with and $b = 0.88(6) $ 
without the penalty term, respectively. The remaining best fit
parameters can be found in \cref{tab:fitParams}. Uncertainties are
estimated from the inverse hessian, rescaled by the variance of the
residuals. 
The analytically predicted result at this $g^2$-value is reproduced
well in the limit $N \rightarrow \infty$, with and without penalty
term.

Furthermore, the results of the extrapolations at all our $g^2$-values
can be found in \cref{fig:mtContAfterGrp}.
Here we also show linear fits of the expected weak coupling form
\begin{equation}
  M(g^2) = M_0 + c \, g^2 
\end{equation}
to the data with and without penalty term.
These allow us to extrapolate to $g^2=0$, i.e. take the weak coupling limit.
Both of these fits agree within the estimated uncertainty with the analytic
prediction plotted in black. At $g^2=0$ we get $M_0=4.008(27)$ with and
$M_0=3.999(3)$ without the penalty term, both of which are compatible with
$M_0=4$.

\begin{table}
  \centering
  \begin{tabular}{c||c|c||c|c}
                 & \multicolumn{2}{c||}{w/o $\hat{H}_{\textrm{penalty}}$} &
    \multicolumn{2}{c}{w. $\hat{H}_{\textrm{penalty}}$}                     \\
    \hline
    $g^2$        & $M^0_{g^2}$                                            &
    $N^0_{g^2}$    & $M^0_{g^2}$                                            &
    $N^0_{g^2}$                                                               \\
    \hline
    $0.0050$     & $3.992(6)$                                             &
    $3.0(4)
    \times 10^2$ & $4.00(5)$                                              &
    $2.97(16)
    \times 10^3$                                                            \\
    $0.0069$     & $3.992(6)$                                             &
    $3.0(4)
    \times 10^2$ & $3.99(5)$                                              &
    $2.96(16)
    \times 10^3$                                                            \\
    $0.0097$     & $3.992(6)$                                             &
    $3.0(4)
    \times 10^2$ & $4.00(5)$                                              &
    $2.88(15)
    \times 10^3$                                                            \\
    $0.0134$     & $3.991(6)$                                             &
    $2.9(4)
    \times 10^2$ & $4.00(4)$                                              &
    $2.81(15)
    \times 10^3$                                                            \\
    $0.0186$     & $3.988(5)$                                             &
    $2.7(4)
    \times 10^2$ & $4.01(4)$                                              &
    $2.61(15)
    \times 10^3$                                                            \\
    $0.0259$     & $3.981(5)$                                             &
    $2.5(4)
    \times 10^2$ & $3.98(4)$                                              &
    $2.67(15)
    \times 10^3$                                                            \\
    $0.0360$     & $3.973(5)$                                             &
    $2.2(4)
    \times 10^2$ & $4.00(4)$                                              &
    $2.41(15)
    \times 10^3$                                                            \\
    $0.0500$     & $3.959(4)$                                             &
    $1.9(3)
    \times 10^2$ & $3.96(4)$                                              &
    $2.55(15)  \times 10^3$
    \\
  \end{tabular}
  \caption{Fit parameters according to \cref{eq:fitInN} for the extrapolation
    to the full gauge group at the eight different couplings.}
  \label{tab:fitParams}
\end{table}

\section{Discussion}

A few of our observations deserve further discussion. The newly constructed
$\hat{Q}^2$ operator, needed for the Gauss law penalty term, seems to work well.
The deviations from the analytic prediction increase, but the
convergence rate seems to be unchanged. It is likely that this effect can be
further reduced by tuning $\kappa$, the prefactor of the penalty term, more
carefully. Penalty terms constructed from the dual Gauss laws found in
Ref.~\cite{Bauer:2023jvw} and Ref.~\cite{PhysRevD.107.074504} would also both contain
this operator. The fact, that it can be used in simulations without major
complications is thus quite reassuring.

Furthermore, considering partitionings with points only distributed in the
vicinity of $\mathbb{1}$ seems to be a valid strategy to approach the weak
coupling limit. When tuning the truncation parameter $\varepsilon_\mathrm{Tr}$
appropriately, the relative deviations of the mass gap were shown to be largely
independent of the coupling at fixed operator dimension. Similar to the
approach presented in Ref.~\cite{Bauer:2023jvw}, we showed that the truncation
parameter $\varepsilon_\mathrm{Tr}$ scales linearly with $g$ at small
couplings.

Simulations using the penalty term typically were more reliable when slightly
increasing $\varepsilon_\mathrm{Tr}$ compared to the simulations with manual
state selection. This is likely because discretisation errors at the boundary
of the truncation have more of an impact when filtering for physical states.
Increasing the size of the sphere cap means that these effects are more
suppressed. One possibility to address this in the future would be to test
whether non-uniform partitionings can be used. Here one would aim for a high
density of points around $\mathbb{1}$ and a decreasing density in the rest of
the group. This should in principle reduce boundary effects and might lead to
more accurate simulations.

Another open question is the convergence rate of the observables.
Reliable extrapolations for $N \to \infty$ were only possible when including
the exponent of convergence as a fit parameter. In our previous
tests~\cite{Jakobs:2023lpp} we found a convergence rate of $N^{-2/3}$
for the 
spectrum of $\sum_c (\hat{L^c})^2$ which would be proportional to the lattice
spacing squared in the partitioning. Here we observe rates of $N^{-0.88(6)}$ and
$N^{-1.00(11)}$ depending on whether a penalty term to enforce Gauss
law is used or not. Currently, we do not have an analytic prediction
of these rates available.

Lastly, it should be mentioned that similar numerical tests have been conducted
in Ref.~\cite{Bauer:2023jvw}. They propose a different discretisation
scheme and
achieve good matching with the analytic prediction at significantly smaller
operator dimension.

\section{Conclusion and Outlook}

In this paper we have shown, that the digitised canonical momentum operators
for the Hamiltonian formulation of an SU$(2)$ gauge theory,
originally proposed in Ref.~\cite{Jakobs:2023lpp}, represent an efficient choice for
simulations at very weak couplings.

In this approach the Hilbert space is digitised by choosing a finite set of
gauge group elements, called a partitioning. The canonical momentum operators
are then approximated by finite element methods. While these operators break
the fundamental commutation relations of the theory, they are local in the
gauge group. We have shown here how to define a penalty term based on
the squared Gauss operator approximated again using finite element
methods to project to physical states of the system.
Given a suitable dual formulation of the Kogut-Susskind
Hamiltonian with a local magnetic term, one can thus use partitionings with
points distributed only around the identity. We show numerically that this
ansatz allows us to approach the weak coupling limit at constant operator
dimensions for a single plaquette system.

For this we first study how to truncate these partitionings at a given coupling
$g^2$. We showed, that the cut-off parameter $\varepsilon_{\mathrm{Tr}}$ as
defined in \cref{eq:partitioningTruncation} scales linearly with $g$,
in agreement with results from Ref.~\cite{Bauer:2023jvw}. By
choosing $\varepsilon_{\mathrm{Tr}}$ optimally for each coupling value, we then
numerically show that the relative deviations of the mass gap of the theory
are independent of the coupling value.

Lastly we have demonstrated that the correct mass gap of the theory is
recovered, when extrapolating first to infinitely fine partitionings and then
to $g^2 \to 0$. This leaves us hopeful that these operators might find
use for the simulation of larger systems at very weak couplings in the future.
However, other approaches like e.g.\ the one from
Ref.~\cite{Bauer:2023jvw} achieve the same, but currently appear to be more
resource efficient for the system investigated here. 
In general though, it is still unclear which approach of the many
available on the market will be most
suitable for the simulation of larger systems with three spatial
dimensions and SU$(3)$ as the gauge group.

\begin{acknowledgments}

  This project was funded by the Deutsche Forschungsgemeinschaft (DFG,
  German Research Foundation) as a project in the
  Sino-German CRC110 and the CRC 1639 NuMeriQS -- project no.\ 511713970.
  This work is supported by the European
  Union’s Horizon Europe Framework Programme (HORIZON) under the ERA Chair
  scheme with grant agreement no. 101087126.
  This work is supported with funds from the Ministry of Science,
  Research and Culture of the State of Brandenburg within the Centre
  for Quantum Technologies and Applications (CQTA).
  \flushright{\includegraphics[width=0.08\textwidth]{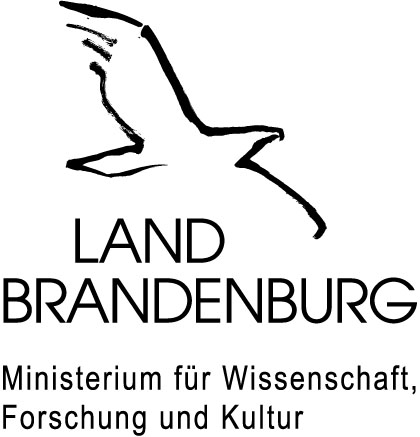}}
\end{acknowledgments}

\bibliography{bibliography}


\end{document}